\documentclass[11pt,a4paper]{article}
\pdfoutput=1
\usepackage[utf8]{inputenc}
\usepackage{jheppub}
\usepackage{slashed}
\usepackage{tikz}
\usepackage{bickham}

\newcommand{\diff}{\mathrm{d}}
\usepackage{slashed}
\usepackage{amssymb}

\usepackage{braket}

\usepackage{graphicx}
\usepackage{wrapfig,enumerate,slashed}

\usepackage{footmisc}
\usepackage{amsmath}
\usepackage{wasysym} 
\usepackage{color}
\usepackage{comment}
\usepackage{ulem}

\DeclareMathAlphabet{\mathbold}{U}{zeur}{b}{n}

\usepackage[labelfont=bf,font={sl,small}]{caption}

\newcommand{\eq}[1]{Eq.~(\ref{#1})}

\renewcommand\[{\left[}
\renewcommand\]{\right]}

\def\beq{\begin{equation}}
\def\eeq{\end{equation}}
\def\bea{\begin{eqnarray}}
\def\eea{\end{eqnarray}}
\def\[{\begin{equation}}
\def\]{\end{equation}}

\begin{document}
\numberwithin{equation}{section}

\title{Probing Poincar\'e Violation}

\author{Rick Gupta$^{1,2}$, Joerg Jaeckel$^{3}$ and Michael Spannowsky$^{1}$}

\affiliation{$^{1}$Institute for Particle Physics Phenomenology, Department of Physics, Durham University, Durham, DH1 3LE, UK}
\affiliation{$^{2}$ Tata Institute of Fundamental Research, Mumbai 400005, India}
\affiliation{$^{3}$Institut f\"ur theoretische Physik, Universit\"at Heidelberg, Philosophenweg 16, 69120 Heidelberg, Germany}

\emailAdd{rsgupta@theory.tifr.res.in}
\emailAdd{jjaeckel@thphys.uni-heidelberg.de}
\emailAdd{michael.spannowsky@durham.ac.uk}

\abstract{
Time and space translation invariance,  giving rise to energy and momentum conservation, are not only amongst the most fundamental but also the most generally accepted symmetry assumptions in physics. It is nevertheless prudent to put such assumptions to experimental and observational tests. In this note, we take the first step in this direction, specifying a simple periodic time dependence that violates time translation invariance in QED, and setting phenomenological constraints on it. In addition to observational and experimental constraints on time varying couplings, we focus on probes of violation of energy conservation such as spontaneous production of photon and electron pairs and the $e \to e \gamma$ process. We discuss similarities and differences to the discussion of time varying fundamental constants and to the case of a light bosonic dark matter field that usually also causes oscillating effects.

}

\preprint{IPPP/22/42}

\maketitle


\section{Introduction}

Poincare invariance is a fundamental assumption in modern particle physics. Indeed particles themselves are defined as irreducible representations of the Poincare group. Direct products of such one particle states then serve as the asymptotic states over which the S-matrix is defined in Quantum Field Theories (QFT)~\cite{Weinberg1}. Moreover, Poincare invariance is also crucial in constructing a consistent theory of gravity \cite{Weinberg3, Weinberg4, Weinberg5}, as conservation of the energy-momentum tensor -- a consequence of Poincare invariance -- is essential for the critical property of general covariance (see for e.g. Chapter 12.3 of  Ref~\cite{Weinberg2}). 

The main goal of this work is to explore how well this assumption of Poincare invariance has been tested and is, therefore, justified from an experimental and observational point of view. In this, we go to the next step in testing fundamental symmetries, building on the seminal work on testing Lorentz symmetry violation by Kostelecky and others~\cite{Colladay:1998fq,Kostelecky:2002hh}.

The distinctive extra feature of Poincare symmetry compared to Lorentz symmetry is spacetime translation invariance. Accordingly, the main aim of the present work is probing spacetime translation violation. For simplicity, we will focus on the more restricted case of time translations. In future work, we aim to investigate the, at least as attractive, case of a violation of space translation symmetry.

The most direct implementation of time translation invariance is a time dependence of fundamental constants, a subject that has already received considerable attention in the literature (see below for a small selection).  
However, from the symmetry point of view, the crucial consequence of a violation of time translation is the possible non-conservation of the Noether current, i.e. energy conservation. We, therefore, pay special attention to this aspect.

Amongst the first to propose spacetime variation of couplings was Dirac, who proposed that the change of some fundamental constants with the age of the universe could be the explanation for large numbers like the ratio of the strength of electromagnetic and gravitational forces~\cite{dirac}; this idea was further developed by Gamow~\cite{gamow}. However, these works did not develop the underlying mechanism for such time variation. In recent times several works have proposed that the spacetime variation of coupling constants arises due to the dynamics of an underlying scalar (or pseudoscalar) field (see Ref.~\cite{review} for a review). These include the so-called `Bekenstein models' developed solely to study such spacetime variation~\cite{Bekenstein, barrow1, barrow2, barrow3},  models utilising extra-dimensional dynamics~\cite{chodos}, string theory inspired models~\cite{damour1, damour2, damour3, damour4}, studies of  spacetime variation of coupling constants within the framework of grand unification,~\cite{HF1,HF2}, models with environmental dependence of coupling constants~\cite{khoury, olive}, models to explain dark energy~\cite{de1,de2,de3,de4} and wave-like dark matter models~\cite{wdm1,wdm2} (we discuss the relation to this particular case more closely below and in Sect.~\ref{sec:compare}). 
Finally, and perhaps especially relevant for our take on the subject, bounds on Poincare violation from the heating of gasses in the context of Causal set theory have been discussed in~\cite{Dowker, Kaloper}.

Before diving into the details, let us briefly spell out the main spirit of our work. Focusing on symmetry and its consequences, compared to most phenomenological studies of time-varying fundamental constants, we keep a closer eye on potential violations of the energy conservation law.
On the other side, and as we will see in more detail, our effective implementation of the Poincare violation is close to the effects expected for wave-like dark matter bosons. However,  compared to those studies, we remain agnostic about the (long-term) dynamics of the system and the restrictions implied by the specific dynamics required for dark matter. Moreover, we remain open to the idea that the time dependence is due to effects that are not directly linked to a new dynamical field/particle.
To remain entirely agnostic, we also make a distinction between constraints that arise directly from physics happening ``today'', i.e. in the last 50 years, and those that rely on processes taking place in the more distant past, where potentially the size but also other features such as the frequency of the Poincare violating effects may have been different.

Let us briefly outline the main steps we want to take. In the following Sect.~\ref{sec:model} we specify the types of Poincare symmetry violation we want to consider, focusing on the QED sector.
Sects.~\ref{sec:delz}~--~\ref{sec:tilm} are used to collect constraints on the time-translation violating parameters specified in Sect.~\ref{sec:model}. As already mentioned, the specific type of violation of time-translation invariance bears some similarities to the case of light bosonic dark matter. We discuss the differences between these situations in Sect.~\ref{sec:compare}. We also briefly compare our approach to tests of dynamical time variation of fundamental constants. Finally, conclusions and an outlook on further possible directions are given in Sect.~\ref{sec:conclusion}.

\section{Translation violating QED}
\label{sec:model}
Since we want to break spacetime symmetries explicitly, we first have to fix a reference frame. We choose the CMB rest frame as our base frame (BF). The violation of spacetime translations is realised via the spacetime dependence of couplings multiplying Lorentz invariant combinations of the field operators.
In this sense, our work is complementary to that of Ref.~\cite{Colladay:1998fq}, where the operators in the lagrangian are not Lorentz scalars, but the couplings are invariant under space and time translations. It is worth emphasising, however, that it is impossible to break only spacetime translations without breaking also the Lorentz group, as the functional dependence of the couplings on space and time would  differ from one frame to another. There is no notion of even local Lorentz invariance, as the spacetime dependence of couplings always implies the existence of a non-vanishing Lorentz vector given by the derivative of the spacetime-dependent couplings. Therefore, we have to choose a BF. 

Keeping terms only up to the dimension 4 level, we obtain the translation violating QED (TVQED) Lagrangian by allowing space-time dependent couplings in the usual QED Lagrangian,
\beq
L_{\rm TVQED}= i \bar{\psi} \slashed{D}_\mu \psi -{\cal m}(x) \bar{\psi}\psi -i \tilde{\cal m}(x)  \bar{\psi}  \gamma^5 \psi- \frac{\cal{Z}(x)}{4} F_{\mu \nu}F^{\mu \nu} -\frac{\tilde{\cal Z}(x)}{4} F_{\mu \nu}\tilde{F}^{\mu \nu} \,.
\label{lagr}
\eeq
The couplings are all real by hermiticity. Here $\psi$ is the electron field, and $F_{\mu \nu}$ is the field strength of the photon. First, this is the most general Lagrangian up to the dimension-4 level. Furthermore, there is no spacetime-dependent coefficient for the fermion kinetic term, say $e^{\kappa (x)}$, as it can be removed completely by a redefinition of the fermion field; the resulting term $(\partial_\mu \kappa) \bar{\psi} \gamma^\mu \psi$ vanishes by integration of parts and the conservation of the electromagnetic current. The third operator above can be rewritten as, 
\beq
i \tilde{\cal m}(x)  \bar{\psi}  \gamma^5 \psi \to -\frac{\partial_{\mu}\tilde{\cal m}(x)}{2m_{e}} \bar{\psi} \gamma^5 \gamma^\mu  \psi,
\label{alternate}
\eeq
again using integration by parts. This alternate way of writing the operator can sometimes be useful. Finally, note that with its space-time dependent coefficient, the last term, $\sim F\tilde{F}$, is no longer a total derivative.\footnote{Note that, even though the TV couplings do not vanish at infinity, the total derivative can be neglected, as the corresponding surface integral is still zero if the other fields vanish sufficiently quickly at infinity. } 

We want to begin our exploration of translation violation (TV) by considering first only a straightforward time variation in our chosen BF.
We leave a study of TV effects that include spatial variations in this frame and more complicated violations of Poincare invariance for future work. 

It will be convenient to decompose these TV couplings into frequency modes in our BF frame, as follows, 
\bea
\delta {\cal Z}(x)&=&Z(x)-1= \sum_{\omega} \delta Z(\omega) \cos \omega t\nonumber\\
\tilde{\cal Z}(x)&=&\sum_{\omega} \tilde{Z}(\omega) \cos \omega t \nonumber\\
\frac{\delta {\mathcal m}(x)}{{m}_e}&=& \frac{\mathcal{m}(x)-{m}_e}{{m}_e}= \sum_\omega \frac{\delta m(\omega)}{m_e} \cos \omega t\nonumber\\
\frac{\tilde{\cal m}(x)}{{m}_e}&=& \sum_\omega \frac{\tilde{m}(\omega)}{m_e} \cos \omega t \,\, .\nonumber\\
\label{defs}
\eea
where $m_e$ is the measured electron mass. As a consequence of our assumption of temporal but no spatial variation, the above couplings can lead to processes that violate energy conservation but always conserve momentum in our BF.  An example of such a process is the spontaneous production of photons or electron-positron pairs from the vacuum; these processes result in the most robust bounds on the above TV couplings at high frequencies.

The couplings $\delta m$ and $\delta Z$ lead, respectively, to variations of the electron mass and the fine structure constant, where that latter is given by
\bea
\frac{\delta \alpha_{em}}{\alpha_{em}}=-\delta \cal{Z} (x).
\eea

In the reference frame of Earth (that is moving relative to our BF), the cosine functions in \eq{defs} get modified as follows, 
\beq
\cos \omega t\to \cos (\gamma\omega (t'+v x'))
\label{spatial}
\eeq
where  $v\sim 10^{-3}$  is the peculiar velocity of the sun in the CMB frame~\cite{peculiar} and  $\gamma^{-1}=\sqrt{1-v^2}\approx 1$.  As we will see, the above spatial variation can result in forces in the direction of the spatial gradient of the above functions.

In general, our bounds on TV depend on the functional dependence of the above couplings on $\omega$. At first sight, at least two interesting limits present themselves: (1) turning on only a single mode at a time in \eq{defs} and (2) the `white noise' limit where the TV couplings,  $\alpha_i^{\rm TV}=\{\delta Z(\omega), \tilde{Z}(\omega), \delta m(\omega), \tilde{m}(\omega)\}$,  have a constant value independent of $\omega$ (at least over a specific frequency range). As a first step, here we will consider the simpler, former possibility as this will already lead to a good first understanding of the nature of the bounds on TV. That said, we expect that the second case also has interesting new features. We hope to do a more general treatment in future work.

\begin{figure}
\noindent \begin{centering}
\hspace{-2em}\includegraphics[scale=0.2]{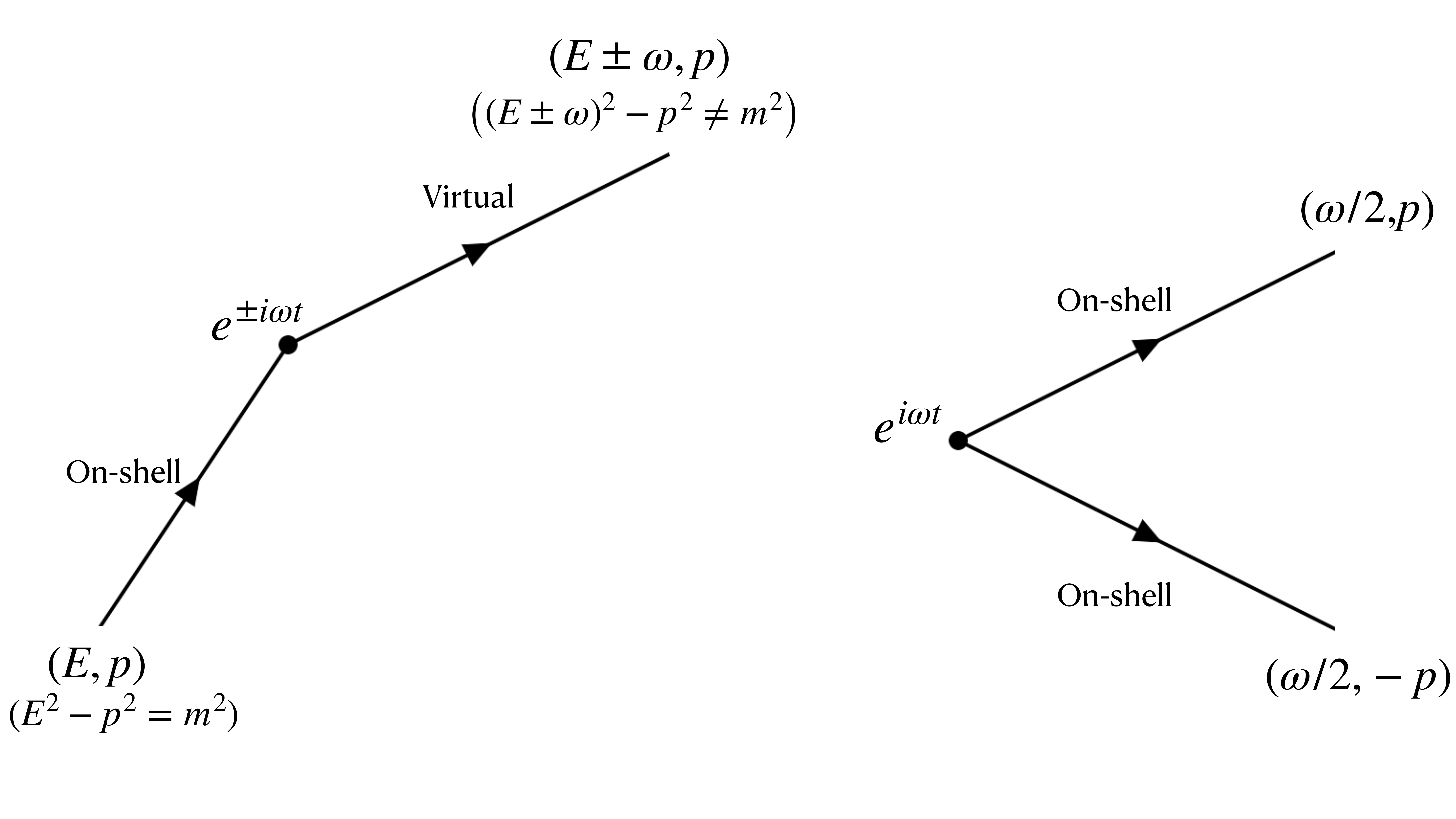}
\par\end{centering}
\protect\caption{We show the effect of a TV insertion on a propagator which cause an increase/decrease
in energy in units of $\omega$. As momentum is conserved in our set-up one of the legs above must be in general
off-shell (left), the only exception being the pair production process (right).}
\label{eviolation}
\end{figure}

\begin{figure}[t]
\noindent \begin{centering}
\hspace{-2em}\includegraphics[scale=0.4]{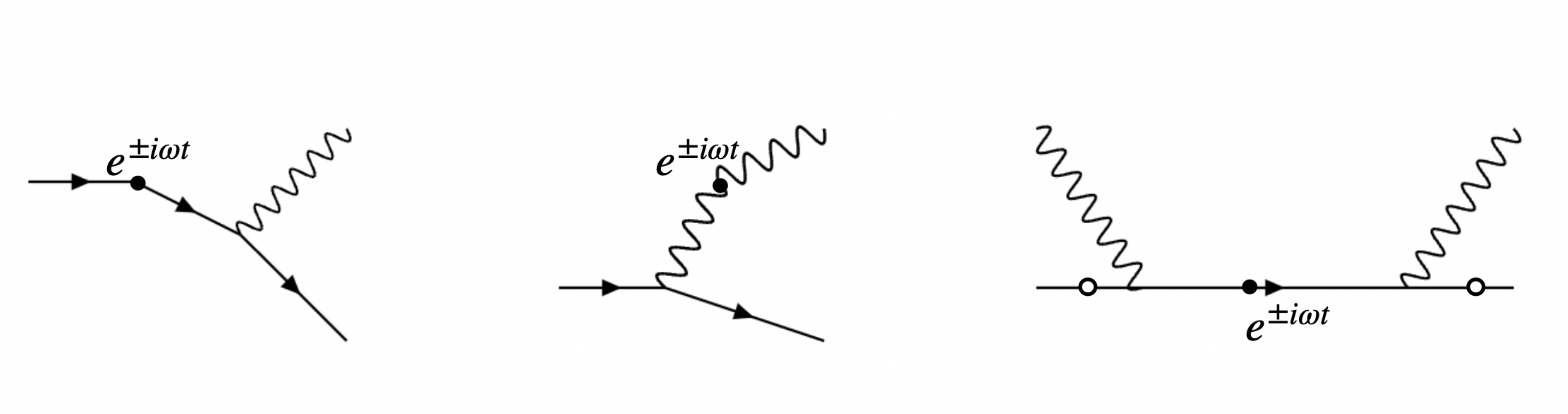}
\par\end{centering}
\protect\caption{{Additional energy conservation violating processes considered in this work. As in Fig.~\ref{eviolation} we show an energy changing insertion into the relevant diagrams. We show only one of the 6 diagrams for the third process; there are two more positions where the TV insertion can be placed (shown by white circles) and  an analogous $u$-channel diagram for each of these 3 diagrams.}}
\label{4figs}
\end{figure}

The TV couplings in \eq{defs}, being quadratic in fields, act as insertions in the Feynman Diagrams as shown in Fig.~\ref{eviolation}. These insertions, however, differ from usual insertions in that they can shift the energy of the propagator by $\pm \omega$. To show this, we consider a single frequency mode and decompose the cosine factor in \eq{defs}, as follows, 
\bea
\alpha_i^{\rm TV}(\omega) \cos~\omega t= \alpha_i^{\rm TV} (\omega) (e^{i \omega t}+e^{-i \omega t}).
\eea
 The $e^{\pm i \omega t}$ factors result in delta functions in the Feynman amplitude,
\bea
\delta(E_f- E_i\pm \omega)
\eea
that allow violation of energy conservation in units of $\omega$, where $E_f$ and $E_i$ are the energies of the two legs of the propagator, as shown in Fig.~\ref{eviolation}.   As we do not consider the violation of spatial translation, the violation of momentum conservation is not allowed in our setup. The requirement of momentum conservation but energy violation by units of $\pm \omega$ means that both the lines in Fig.~\ref{eviolation} can usually not be on-shell. A situation where this is nevertheless possible is pair production from ``vacuum'', which preserves momentum conservation but violates energy conservation as shown in the right-hand part of Fig.~\ref{eviolation}. In all other instances, one of the two lines connected by the insertion must be virtual. 

In this work, we will focus on probing the simplest processes in TVQED that violate energy conservation. At the zeroeth order in $\alpha_{em}$, the only energy conservation violating processes are the pair production of photons and electrons. These processes, however, do not provide bounds below the kinematic threshold, $\omega < 2 m_e$ and $\omega < 2 m_\gamma$, respectively, where $m_\gamma$ is the plasma mass of the photon. To obtain sensitivity in the $\omega < 2 m_e$ regions for the TV couplings $\delta m(\omega), \tilde{m}(\omega)$, we will also study the process with a single virtual electron, i.e. $e \to e \gamma$, which has an ${\cal O}(\alpha_{em})$ cross-section. Even this process has a kinematic threshold $\omega >m_\gamma$, in situations where the photon has a plasma mass. To overcome this, we will also study energy conservation violation in the $e \gamma \to e \gamma$ process. These processes are shown in Fig.~\ref{4figs}. Finally, we will also consider bounds from the analogue of the photoelectric effect, where the emission of electrons takes place due to the TV couplings instead of a photon field.

In the following sections, we will present bounds on the four TVQED couplings arising from the violation of energy conservation (VEC) in these processes. These are then combined with observational and experimental probes of the time variation of couplings constants and results from wave-like dark matter experiments. For some cases, we also consider the constraints from effectively Lorentz violating effects. While the VEC bounds turn out to be the most potent bounds for high-frequency modes, their region of applicability is limited by the kinematic considerations outlined above. The VEC probes are thus complementary to other observational probes that are powerful at lower frequencies but unable to constrain high-frequency modes.

\section{Constraints on $\delta {\cal Z} (x)$}
\label{sec:delz}
We begin with a discussion of the bounds on $Z(\omega)$, defined in \eq{defs}, as a function of the frequency $\omega$. There are three main categories of constraints on this coupling. At high frequencies, the dominant limit is from probes of energy conservation violation, whereas at lower frequencies, the bounds arise mainly from wave-like dark matter experiments and from testing the variation of the fine structure constant. 

\subsection{Bounds from the violation of energy conservation}
\subsubsection{Spontaneous photon production}
\label{delzpp}
{As indicated in Fig.~\ref{eviolation}, the time dependence of  $\delta {\cal Z}$ shown in \eq{defs} leads to spontaneous production of photon pairs which lead to the most powerful bounds on $\delta Z (\omega)$ at high frequencies. The amplitude for spontaneous photon production is,
\bea
M_{n\gamma\to (n+2)\gamma}\!\!&=&\!\!\langle -i(N_{k_1,\epsilon_1}+1),(N_{k_2,\epsilon_2}+1)|\int d^4 x ~\frac{\delta {\cal Z} (x)}{4}F_{\mu \nu}F^{\mu \nu}|N_{k_1,\epsilon_1},N_{k_2,\epsilon_2}\rangle\nonumber\\
\!\!&=&\!\!-i\sum_\omega(N_{{k_1}}+1)(N_{{k_2}}+1)\delta(\omega-E_1-E_2)\delta^3(\mathbf{k_1}+\mathbf{k_2})\frac{ \delta Z(\omega)}{2}V(k_1,k_2,\epsilon_1,\epsilon_2),\nonumber \\
\label{amp1}
\eea
whereas, for  the reverse process we have, 
\bea
M_{(n+2)\gamma \to n\gamma}&=&-i\langle N_{k_1,\epsilon_1},N_{k_2,\epsilon_2}|\int d^4 x ~\frac{\delta {\cal Z} (x)}{4}F_{\mu \nu}F^{\mu \nu}|(N_{k_1,\epsilon_1}+1),(N_{k_2,\epsilon_2}+1).\rangle\nonumber\\
&=&-i\sum_\omega N_{{k_1}}N_{{k_2}}\delta(\omega-E_1-E_2)\delta^3(\mathbf{k_1}+\mathbf{k_2})\frac{ \delta Z(\omega)}{2}V(k_1,k_2,\epsilon_1,\epsilon_2).\nonumber \\
\label{amp2}
\eea
Here $k_i=(E_i, \mathbf{k_i})$ and $\epsilon_i$ are the four-momentum and polarisation four-vectors and,
\bea
V(k_1,k_2,\epsilon_1,\epsilon_2)=\left((k_1.k_2)(\epsilon_1.\epsilon_2)-(\epsilon_1.k_2)(\epsilon_2. k_1)\right).
\eea
For a given frequency mode in \eq{defs}, the growth of the photon number density for a given frequency  is then given by\footnote{This can be obtained by adapting the standard computation of the partial  width from a decay amplitude~\cite{peskin}. Note that if $\delta {\cal Z}=g_{\phi \gamma} \langle \phi \rangle$ arises from an oscillating scalar, $\phi$, our expression gives the correct result $\dot{n}_{\gamma}=2 n_\phi\Gamma_{\phi \to \gamma \gamma}$ with $\omega=m_\phi$, $n_\phi=\frac{1}{2}m_\phi \langle \phi \rangle^2$ and   $\Gamma_{\phi \to \gamma \gamma}=g^2_{\phi \gamma} m^3_\phi \beta_\gamma/64 \pi$. See also, Ref.~\cite{surprise}, where the equivalence of this approach to that using the classical equation of motion has been discussed.},
\bea
\dot{n}_\gamma(\omega)&=& \int (2 \pi)^4  \Bigg(|M_{n\gamma\to (n+2)\gamma}|^2-|M_{(n+2)\gamma\to n\gamma}|^2\Bigg)\frac{d^3 k_1}{(2 \pi)^3 2E_1}  \frac{d^3 k_2}{(2 \pi)^3 2E_2}\, ,\nonumber\\
\label{ngdot}
\eea
which finally gives (see also~\cite{Baumann}),
 \bea
\dot{n}_\gamma(\omega)&=&(2 N_k+1) \frac{ (\delta Z(\omega))^2 \omega^4 \beta_\gamma}{64 \pi},
\label{mel}
 \eea
where $\beta_\gamma=\sqrt{1-4 m_\gamma^2/\omega^2}$ and $m_\gamma$, the thermal mass of the photon,  has a current value  of about $10^{-14}$ eV~\cite{raffelt, braaten, surprise}. 
$N_{k}$ denotes the occupation number of the photons with energy, $\omega/2$, and magnitude of  momentum, $k=\beta \omega/2$, assuming an isotropic photon distribution. 

We will now discuss two different regimes: (1) the Bose-enhanced regime, $N_k \gg 1$, where the prefactor $(2 N_k+1)$ results in exponential photon production, and (2) the perturbative regime, $N_k \sim 1$.

\paragraph{Bose-enhanced regime}

First, let us ignore the effect of the expansion of the universe. In the $N_k \gg 1$ limit we can  rewrite \eq{mel} by substituting for the occupation number (cf., e.g.~\cite{surprise}), 
\bea
N_k \sim \frac{n_\gamma / 2}{4 \pi k_\star^2 \mathop{\delta k}/(2 \pi)^3}
\eea
and using the expression for   ${\delta k}$, the width of the   photon frequency range (see again, e.g., Ref.~\cite{surprise}),  
\begin{equation}
\label{eq:width}
\delta k \sim \delta Z(\omega) \omega/2.
\end{equation} 
We then obtain from \eq{mel}, 
\bea
\dot{n}_\gamma = 2 \eta_k n_\gamma
\eea
with,
\begin{equation}
\label{eq:growth}
\eta_{k}\sim \delta Z(\omega) \omega.
\end{equation}
We refer to Ref.~\cite{surprise} for a detailed discussion of the correspondence between the approach using \eq{mel} to derive this exponential growth and the classical approach using the equation of motion.  

The expansion of the universe eventually stops this exponential growth~\cite{surprise} as it redshifts the produced photons away from the above resonance band in a timescale given by, 
\bea
 {\mathop{\Delta t}}_{\rm redshift} \sim \frac{\mathop{\delta k}}{k_\star} \frac{1}{H}
 \label{1byh}
 \eea
where $k_\star=\omega/2$. To obtain the region excluded due to photon production,  we will require that the energy density of photons, 
\bea
\rho_\gamma(t)= \frac{1}{\pi^2} \int \mathop{\diff k}k^2 \omega_k N_k(t) &\simeq& \frac{1}{\pi^2}\, k_\star^3 \mathop{\delta k} \sqrt{\frac{\pi}{\eta_{k}t}}\,N_k (t)
\label{rhog}
\eea
 with, 
\bea
N_k (t)&=& N_{k_\star}^0 (e^{2 \eta_k t}-1)
\eea
surpasses either the critical density, $H^2 M_{pl}^2$, or the constraints from Extragalactic Background Light (EBL)~\cite{ebl}. Note that we have subtracted the zero-point energy above, which corresponds to an occupation number $N_{k_\star}^0=1/2$ in each state.

To derive the bounds from EBL, we have adapted the calculation of  Ref.~\cite{masso}. The photons produced in a time $\delta t$,
\bea
\delta n_\gamma=\eta n_\gamma \delta t.
\label{difral}
 \eea
with $n_\gamma=\rho_\gamma/(\omega/2)$ can be related to the observed spectrum using $\delta t=H^{-1}d E/E$, 
 \bea
 E\frac{d^2F_n}{dE d\Omega}=-\lambda\frac{d^2F_n}{d\lambda d\Omega}=\frac{1}{4 \pi}n_\gamma(\omega)\frac{\eta}{H}.
 \eea
As we are interested in the bounds on TV today, we have assumed that the redshift, $z\sim 1$. Here $I_\lambda=\frac{d^2F_n}{d\lambda d\Omega}$ is the observed EBL spectrum from Ref.~\cite{ebl}. Then, comparing the produced photons with the data, we can find the number of e-folds of exponential growth, denoted by $\xi$, that saturates the EBL bound~\footnote{To be conservative, we always take $\xi\geq 1$.}. This bound can then be translated to a bound on $Z(\omega)$ using Eq.~\eqref{pp}.
To do this, we require that the time period for sufficient exponential growth,  $\xi/\eta_k$, is smaller than ${\Delta t}_{\rm redshift}$, i.e. $\xi/\eta_k <{\mathop{\Delta t}}_{\rm redshift}$,\footnote{Note that these bounds are conservative because photon production continues even after the exponential phase stops. } which gives, 
 \bea
\delta{Z}(\omega) &<& \sqrt{\frac{2 \xi   H}{\omega}}.
\label{pp}
\eea
While this constraint must always be satisfied,  the most robust bounds can be derived by imposing this constraint today. 
Note that the time period for stopping the exponential phase in \eq{1byh} is much smaller than one Hubble time by a factor $ {\mathop{\delta k}}/{k_\star}=\delta Z(\omega)$. Moreover, the value of ${\mathop{\Delta t}}_{\rm redshift}$ is smaller than 50 years so that this bound indeed probes violation of time translation invariance today and does not assume the presence of these effects in the distant past.
The corresponding excluded region is shown in magenta in  Fig.~\ref{photon}.

\paragraph{Perturbative regime}
From  \eq{rhog}, one can see that assuming $\rho_\gamma<\rho_c$,  the critical density, the classical limit $N_k \gg 1$ will not hold at higher frequencies. At these frequencies, the density of states is so large that photon production surpasses observational bounds even with small occupation numbers. In this regime, we can use \eq{mel},  with $N_k \sim 1$, and demand that the amount of electromagnetic energy produced in a time period, $\Delta t= 50$ years (5 Gyr), is smaller than the critical density,
\bea
\dot{n}_\gamma \omega \Delta t <\rho_c.
\eea
The resulting bounds are shown in orange magenta (hatched magenta) in Fig.~\ref{photon}. 

 \begin{figure}
\noindent \begin{centering}
\hspace{-2em}\includegraphics[scale=0.5]{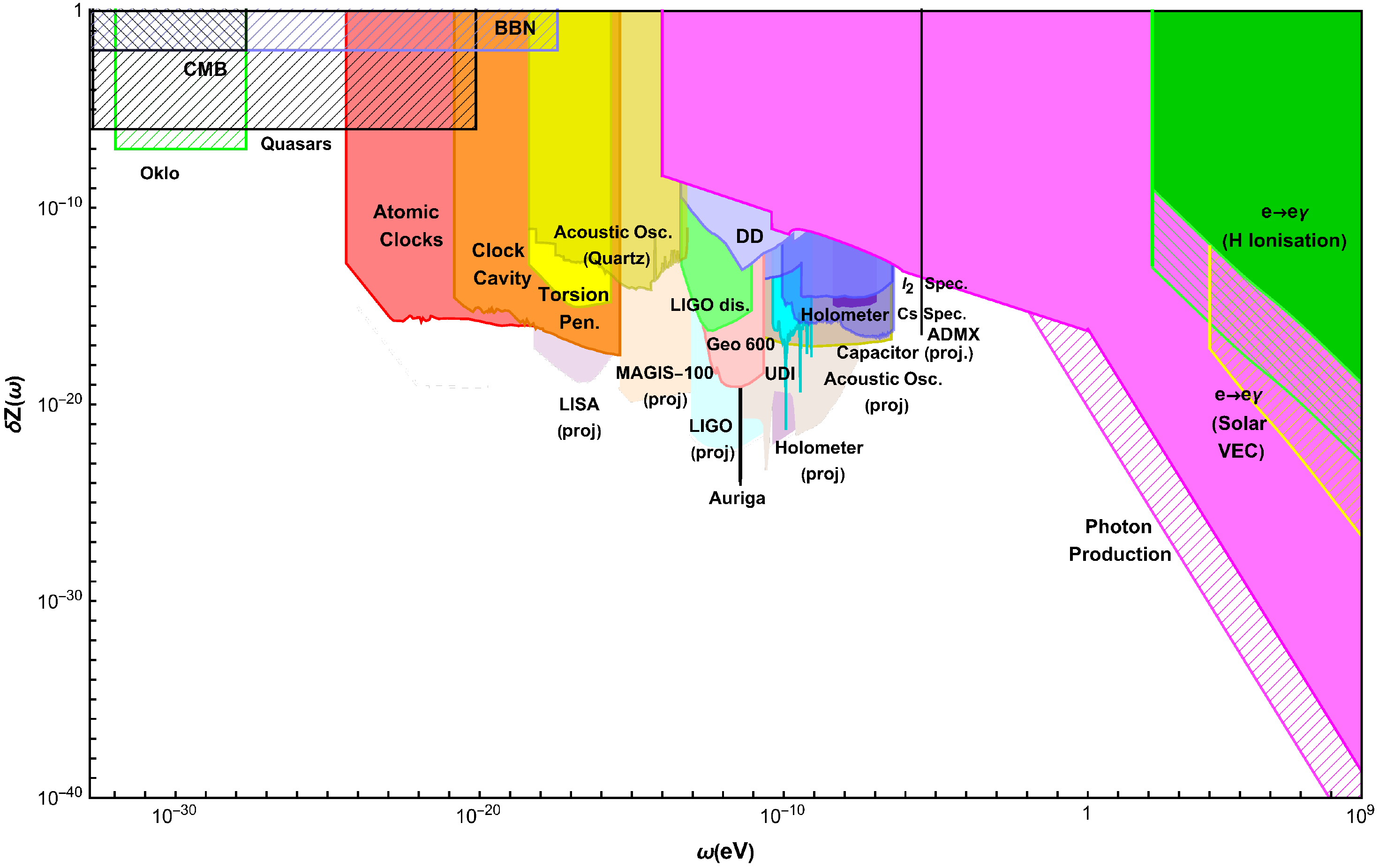}
\par\end{centering}
\protect\caption{Bounds on ${Z}(\omega)$ from VEC, observational and experimental probes. Solid areas are constraints where the TV has to be present only in recent times ($\sim 50$~years) whereas hatched areas require its presence on cosmological time-scales. See the text for more details.} 
\label{photon}
\end{figure}

\subsubsection{$e\to e \gamma$ process}
\label{eeg1}
The $e\to e \gamma$ process can be induced by a $\delta Z (\omega)$ insertion into the diagram for the usually forbidden $e\to e\gamma$ decay, as shown in  Fig.~\ref{4figs}.

\paragraph{Ionisation of Hydrogen}

For $\omega \gg \Delta E_B$, the binding energy of molecular  Hydrogen, the $e\to e \gamma$ process would lead to the ionisation of Hydrogen. The typical ionisation would be given by the inverse of the rate of the $e\to e \gamma$ process. Requiring this ionisation time of Hydrogen to be greater than 50~years (the age of the universe) gives the bound shown by the region shaded green (hatched green) in Fig.~\ref{photon}.

\paragraph{Violation of Energy conservation in the sun}
The $e\to e \gamma$ process will also lead to VEC  in the sun and other astrophysical systems. To compute the rate of energy production/depletion, we perform a thermal average of the  $e (p_1)\to e(p_2) \gamma (p_3)$ rate and subtract the rate of the inverse process. We finally obtain, 
\begin{equation}
\frac{dE}{dt}=\omega N_e \int \prod_{i=1}^3 \frac{d^3 p_i}{2 \pi E_i}
\left(|{  M}_{e \to e \gamma}|^2f_{\rm MB}(p_1)-|{  M}_{e\gamma \to e}|^2 f_{\rm MB}(p_2) f_{\rm BE}(p_3)\right)2 \pi^4 {\delta^4\left(\sum_{i}p_{i}\right)}
\label{dedt}
\end{equation}
where $N_e$ is the number of electrons,
\bea
f_{\rm MB}(p)&=&\frac{1}{(2 \pi m_e T)^{3/2}}\exp\left(-\frac{p^2}{2 m_e T}\right)\nonumber\\
f_{\rm BE}(p)&=&\frac{1}{(2 \pi)^3}\frac{1}{\exp{(-p/T)}-1}.
\eea
The factor inside the parenthesis is positive, so the net effect is heating up the sun.  

We use the values from Ref.~\cite{solarmodel,RedondoSolar} for the solar temperature and density profiles. To get a conservative bound, we consider only the sun's central part with a radius of 20$\%$ of the entire solar radius. In this region, we can take the temperature, $T=1$ keV,  the density, $\rho=30$ g/cm$^3$ and the number of mass units per electron, $\mu_e=1.5$. The total number of  electrons in this region is then given by,
\bea
N_e=\frac{\rho N_{\rm Av}}{\mu_e} \frac{4 \pi (0.2 R_s)^3}{3}
\eea
where $N_{\rm Av}$ is the Avogadro number and $R_s$ is the radius of the sun. Requiring that the energy production does not change the sun's luminosity by more than $10\%$, we obtain the bounds shown in hatched yellow lines in  Fig.~\ref{photon}. 

Plasma effects play an essential role in the process we are considering. There is a suppression in the rate if the momentum exchange via the virtual photon  is less than the momentum scale, $k_D\approx 10$ keV,  corresponding to the Debye screening length~\cite{Raffelt1986}. For frequencies, $\omega\ll m_{e}$, the momentum transfer in the virtual photon is, ${\cal O}(\omega)$,  so the bound will be suppressed for $\omega < k_D$. As this is not the most stringent bound in this region of the parameter space, we simply truncate our bound for,  $\omega\leq k_D$, instead of computing the suppression more carefully.

The bounds obtained from the $e \to e \gamma$ process are weaker than those obtained from photon-pair production because the former process is $\alpha_{em}$ suppressed concerning the latter.

\subsection{Gradient Forces}
\label{gradForce}

A spatial variation of the fine structure constant or the electron mass induces a force  given by,
\beq
\textbf{F}_{spatial}=-\nabla M_{t}(\alpha_{em}(x),m(x))
\eeq
on a test body of mass, $M_t$. This is because a space-time-dependent mass is a potential for the test mass. 

Per assumption, we do not consider spatial variations of the fundamental constants in the BF. However, as the Earth moves relative to the BF, we will have a spatial variation according to Eq.~\eqref{spatial}. Similar to the wave-like dark matter discussed in~\cite{gkr}, we then have an oscillating (in time) force aligned with the direction of movement through the BF.

The mass of each atom in the test body receives contributions from the electron mass and the electrostatic energy of the proton distribution  within the nucleus; these  are respectively given by the first and the second term below~\cite{Grote:2019uvn}, 
\beq
\delta M_{atom}\approx Z_A \delta{\cal m}(x') + \frac{Z_A^2 a_C}{A^{1/3}}\delta {\cal Z}(x')
\label{delM}
\eeq
where $Z_A$ is the atomic number, $A$ is the total number of nucleons. The primed coordinates in the above equation represent the earth reference frame where   $\delta{\cal m}$ and $\delta {\cal Z}$ obtain a spatial dependence as shown in \eq{spatial}.   The dependence on $\alpha_{em}$  in the above equation arises via $a_C=0.7$ MeV. 

\paragraph{Torsion Pendulums} As proposed in Ref.~\cite{gkr}, the gradient force described above can be probed by torsion pendulums used in fifth force experiments. The induced gradient force oscillates at a frequency different from the pendulum's natural frequency and points in a direction different from the vertical static forces due to the earth. This allows these gradient forces to be probed very sensitively by Torsion pendulum experiments. The Washington group has recently presented its first results constraining such a signal in Ref~\cite{wash}. The bounds on the variation of $\alpha_{em}$ arise from the second term in \eq{delM} where the values of $Z_A$ and $A$ have been appropriately chosen for the Be-Al composition dipole used in Ref.~\cite{wash}; the resulting bounds are shown in Fig.~\ref{photon}. In Fig.~\ref{photon}, we also show the projection in Ref.~\cite{gkr} for a future improved torsion pendulum set-up.

\paragraph{LIGO}  Gradient forces would also lead to a time-dependent centre of mass displacement of the test masses in interferometers like LIGO. The resulting bounds were computed in Ref.~\cite{Guo:2019ker,Grote:2019uvn} and reproduced in Fig.~\ref{photon}.

\subsection{Haloscopes}

Haloscope experiments use cavity resonators to enhance electromagnetic signals by converting ALP dark matter to photons (see Sect.~\ref{haloscopeSec}). The same principle can be utilised to detect the effect of  $\delta Z(\omega)$, which can effectively arise from oscillating scalar dark matter. In the scalar case, however, the roles of electric and magnetic fields are reversed so that the usual configuration used in haloscopes with a uniform magnetic field is not sensitive to scalar dark matter~\cite{admxS}. As the magnetic field is not uniform in experimental situations, there is still some sensitivity to $\delta Z(\omega)$~\cite{admxS}. One can, for instance, translate existing bounds from the ADMX experiment to obtain bounds on  $\delta Z(\omega)$; we show the bounds obtained in Ref.~\cite{admxS} by the vertical black line Fig.~\ref{photon}. However, much better sensitivity can be obtained by adapting the experimental set-up to this scenario, as explained in Ref.~\cite{admxS}. They show, for instance, that a capacitor with a uniform electric field can provide strong bounds on $\delta Z(\omega)$ for an extensive range of frequencies; the projected bounds obtained are shown in Fig.~\ref{photon}.

\subsection{Time variation of the fine structure constant}

Powerful bounds at low frequencies arise from probing the variation of the fine-structure constant induced by $\delta {\cal Z}(x)$. The two primary ways of probing such a variation are: (1) a comparison of the present value of  $\alpha_{em}$ with that at earlier times and (2) experimental measurements. In the following, we quickly recap some of the most relevant constraints for our setup.

\subsubsection{Comparison with value of $\alpha_{em}$ at earlier times}
\label{photoncomp}

\paragraph {BBN, CMB and quasars} The earliest observations that can be used to infer a value for $\alpha_{em}$ are those related to Big bang nucleosynthesis. BBN observables constrain any variation in $\alpha_{em}$ to be less than per cent level~\cite{photonbbn}. This constraint implies a bound in the  frequency range  $10^{-33}{\rm eV}\lesssim \omega \lesssim10^{-17}$ eV . Below this range, the time period is larger than the time between BBN and the present day. For frequencies higher than the range shown, the time period is much smaller than 3 minutes, the BBN timescale. The bound, $\delta \alpha_{em}/\alpha_{em}\lesssim 0.01$ arise from CMB observables~\cite{cmb} in the frequency range $10^{-33}{\rm eV}\lesssim \omega \lesssim10^{-28}$ eV. The frequency range corresponds to oscillation periods smaller than the time between recombination and today but larger than the recombination timescale $\sim 100$ kyr.

Even stronger bounds can be inferred from observing absorption lines in the spectra of distant quasars~\cite{Webb99,Webb00,Webb01,Chand, Srianand}. Many such observations have been made by the Keck/HIRES and VLT/UVES telescopes as reviewed in Ref.~\cite{review}. These constrain variations of fundamental constraints at the parts per million level, i.e. $\delta \alpha_{em}/\alpha_{em},\delta m/m \lesssim 10^{-6}$,  as shown in Fig.~\ref{photon} and \ref{gradient}. These observations lie in the redshift range $0.2<z<6.4$, so there are no bounds if the time variation scale is greater than 10 Gyr. On the other hand, time variation at a scale faster than a day also cannot be bounded by this method as the observation time required for a sufficiently large signal-to-noise ratio is of this order~\cite{Srianand2}. 

\paragraph{The Oklo bound} 
Bounds stronger than the astrophysical and cosmological ones described above arise from a comparison with $\alpha_{em}$ inferred from the Oklo phenomenon as first suggested by Ref.~\cite{Shl1,Shl2}. The Oklo phenomenon was a natural fission reactor operated around 2 billion years ago for $\sim$100,000 years. The bound arises from the tiny amounts of the isotope Sm$^{149}$, a product of U$^{235}$ fission, that were found. This is consistent with the effect of neutron flux on  Sm$^{149}$, which has an enormous cross-section for absorbing a free neutron. This enormous cross-section arises because of a resonance state just above zero to which  Sm$^{149}$  can be excited by the capture of a free neutron. It was shown in Ref.~\cite{damour}  that a value of $\delta \alpha_{em}/\alpha_{em}$ larger than $10^{-7}$ shifts the resonance energy level,  and thus the neutron capture cross-section, far too much to be compatible with the observed Sm$^{149}$ abundance. We show the resulting constraints in Fig.~\ref{photon}.

\subsubsection{Experimental probes of a variation of $\alpha_{em}$}
\label{photonexp}
\paragraph {Atomic Clocks} The spacetime dependence of the fine-structure constant can be constrained using many different techniques~\cite{feeble}. Atomic clocks provide powerful bounds in the frequency range, $10^{-25}$ eV$-10^{-15}$ eV, corresponding to frequencies below 1 Hz. These arise due to a variation of atomic transition frequencies due to the underlying modulation of fundamental constants like the fine structure constant and the electron mass. The region shaded red in Fig.~\ref{photon} combines the bound derived in Ref.~\cite{Hees:2016gop} from a measurement of the ratio of hyperfine transition frequencies of Rb and Cs and the bound in Ref.~\cite{tilburg} from a spectroscopic analysis in two isotopes of dysprosium. 

Beyond that, recent developments have pushed the reach of atomic and molecular spectroscopy to probe $\alpha_{em}$ variations to frequencies as high as 100 MHz. We show the bounds dynamic decoupling (DD)~\cite{DD}, Cs transitions~\cite{budkerJ} (see also Ref.~\cite{antypas}) and transitions in $I_2$ molecules~\cite{budkerN}  in different shades of blue.

\paragraph {Clock-cavity comparison} Strong bounds can be derived from the fine structure constant variation by comparing cavities' frequency with transition frequencies in atomic clocks. For example, in Fig.~\ref{photon}, we show in orange the bounds from such `clock-cavity' comparisons in Ref.~\cite{Kennedy:2020bac}. 

 \paragraph {Atomic and optical interferometry} At somewhat higher frequencies, strong bounds can be derived by probing the effect of fine structure constant variation on existing optical interferometers such as the ones in GEO-600, LIGO, the Fermilab holometer and LISA~\cite{Grote:2019uvn}. The variation of the fine structure constant can affect these interferometers in three ways. The first way is by changing the length of the interferometer arms owing to a variation in the Bohr radius. Secondly, it can cause a variation in the refractive index inside the interferometer. Finally, and as already discussed, the spatial gradient of this variation in the earth frame (see Sect.~\ref{gradForce})  can lead to a `force' on the test masses inside the interferometers. The bounds from Geo-600~\cite{geo} and the Fermilab holometer~\cite{holo} are shown in Fig.~\ref{photon} in pink and purple, respectively. We also show the projections for  LIGO, the Fermilab holometer and LISA from Ref.~\cite{Grote:2019uvn}   in Fig.~\ref{photon}.

In Fig.~\ref{photon}, we also show in cyan the bounds obtained in Ref.~\cite{Savalle}  by probing the variation of the frequency of an optical cavity using an unequal delay interferometer (UDI). 

 Bounds can also be derived from atom interferometry, utilising atomic waves~\cite{gkr,aint2}. In Fig.~\ref{photon}, we show the atom interferometry projections from the proposed MAGIS-100 experiment~\cite{magis1,magis2}. 
 
 \paragraph {Acoustic Oscillations}  It was proposed in Ref.~\cite{dmsound1} that the time variation of fundamental constants can lead to acoustic oscillations in the size of solids. The bounds from the  AURIGA experiment~\cite{auriga} and an experiment probing this effect using a quartz crystal~\cite{dmsound3} are shown in black and light brown in Fig.~\ref{photon}. We also show the original projections 
of Ref.~\cite{dmsound1} for a Cu-Si system in Fig.~\ref{photon}.

\section{Constraints on $\tilde{\cal Z}(x)$}
\label{sec:tilz}
While at first sight, the $\tilde{\cal Z}(x)$ coupling looks very similar to $\delta{\cal Z}(x)$ its pseudoscalar nature eliminates the bounds that result from (oscillating) macroscopic forces as well as those from shifts in energy levels (e.g. spectroscopy) at leading order. What remains are limits from the spontaneous particle production and much stronger constraints from experiments searching for axion dark matter (which is targeted at pseudoscalar effects).

\subsection{Bounds from violation of energy conservation}
\subsubsection{Spontaneous photon production}
Just as in Sect.~\ref{delzpp},  the time dependence of   $\tilde{\cal Z}$  leads to the spontaneous production of photon pairs. \eq{amp1} and \eq{amp2} still hold but with the following replacements,
\bea
&&\frac{\delta {\cal Z} (x)}{4}F_{\mu \nu}F^{\mu \nu} \to \frac{\tilde{\cal Z} (x)}{4}F_{\mu \nu}\tilde{F}^{\mu \nu}~~~\delta {Z} (\omega) \to \tilde{Z}(\omega),
\nonumber\\
&&V(k_1,k_2,\epsilon_1,\epsilon_2) \to \tilde{V}(k_1,k_2,\epsilon_1,\epsilon_2)=\left(\epsilon_{\alpha\beta \mu \nu}k_1^{\alpha}\epsilon_2^{\beta}k_2^{\mu}\epsilon_2^{\nu}\right).
\eea
For the growth of the photon density, we then obtain, 
 \bea
 \dot{n}_\gamma(\omega)&=&(2 N_k+1) \frac{ (\tilde{Z}(\omega))^2 \omega^4 \beta_\gamma}{64 \pi }.
\label{mel2}
 \eea
which is identical to   \eq{mel} up to the replacements above. The discussion in Sect.~\ref{delzpp}, thus holds completely, and we again arrive at the same conditions for both the Bose-enhanced and perturbative regimes, 
\bea
&&\tilde{Z}(\omega) < \sqrt{\frac{2 \xi   H}{\omega}}\nonumber\\
&&\frac{\dot{n}_\gamma \omega}{H}<\rho_c
\label{ppp}
\eea
for the allowed region in our parameter space. We show these in orange in Fig.~\ref{photon2}. Once again, $\xi$ is the number of e-folds of the exponential production required to surpass either the critical density or the EBL bounds discussed in Ref.~\ref{delzpp}.
\subsubsection{$e\to e \gamma$ process}
\label{eegzt}
The $e\to e \gamma$ process can also be induced by a $\tilde {Z} (\omega)$ insertion in a Primakoff-like process (see Fig.~\ref{4figs}). Again we proceed as described in Sect.~\ref{eviolation} to obtain the amplitude. This agrees with the $e \gamma \to e \tilde{\phi}$ amplitude in Ref~\cite{cpn}, adapted to our case. Proceeding as in Sect.~\ref{eeg1}, we again get bounds shown in Fig.~\ref{photon2} by demanding that the ionisation time of hydrogen is less than 50 years (solid green) or the age of the universe (hatched green). The limit from an increase of the solar luminosity over a long period of time is shown as the hatched yellow region.

 \begin{figure}
\noindent \begin{centering}
\hspace{-2em}\includegraphics[scale=0.5]{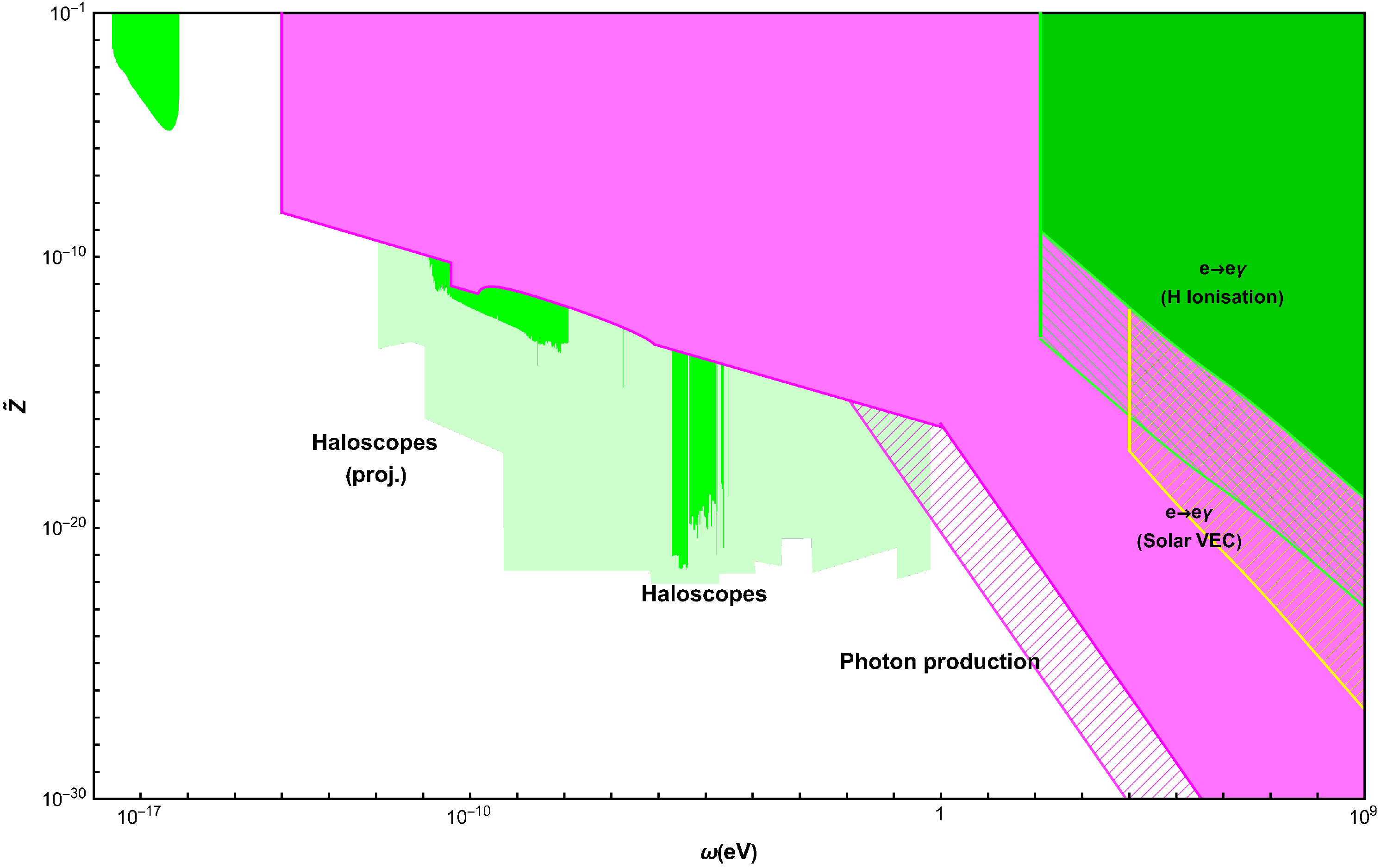}
\par\end{centering}
\protect\caption{Bounds on $\tilde{Z}(\omega)$ from VEC, observational and experimental probes. Solid areas are constraints where the TV has to be present only in recent times ($\sim 50$~years), whereas hatched areas require its presence on cosmological time scales. See the text for more details.}
\label{photon2}
\end{figure}

\subsection{Experimental bounds from haloscopes}
\label{haloscopeSec}
Haloscopes are experiments designed to detect axion (or ALP)  dark matter. In the presence of the oscillating axion background, the Maxwell equations of electromagnetism are modified such that a magnetic field induces an oscillating electric field. The same effect also exists in the presence of a time-dependent $\tilde{Z}$. In conventional haloscopes, this electrical field is enhanced in a microwave cavity. Different groups have adapted and modified this basic idea to gain sensitivity in different mass ranges. An updated compilation of bounds and sensitivity projections for different haloscopes can be found in Ref.~\cite{web}. We show the bounds on $\tilde{Z}(\omega)$ from Refs.~\cite{abrarun2, admx10,admx18,admxside,admxslic,admx21,base,capp1,capp2,capp3,grahal,haystac1,haystac2,organ,quax19,quax20,rades,rbf,shaft,supermag,uf,upload}
 in dark green, whereas the projected sensitivities from Refs.~\cite{abraproj, abmc,admxproj,aligoproj,breadproj,brassproj, danceproj, dmradio1, dmradio2, klash, lamppost, madmax,organ,srf,toorad, wisp} in light green in Fig.~\ref{photon2}. We again refer to Ref.~\cite{web} for detailed individual labelling of the different experiments.


\section{Constraints on $\delta {\cal m}(x)$}
\label{sec:delm}
\subsection{Bounds from violation of energy conservation}

\subsubsection{Spontaneous production of electron-positron pair}
\label{eepp}

In principle, the calculation of the spontaneous pair creation, in this case, electron-positron pairs, proceeds similarly to photon pair production. We will, therefore, mainly focus on the difference that arises because electrons are affected by the Pauli principle.

The time dependence in  $\delta m$ shown in \eq{defs} leads to the spontaneous production of electron-positron  pairs, with the amplitude given by,
\bea
M_{0\to ee}&=&-i\langle \bar{\psi}(k_1,s_1) \psi(k_2,s_2) | \int d^4 x ~\delta{\cal m}(x)~\bar{\psi}(x)\psi(x)|0\rangle
\nonumber\\&=&-i\sum_\omega (2 \pi)^4 \delta(\omega-E_1-E_2)\delta^3(\mathbf{k_1}+\mathbf{k_2})   \bar{u}_{s1}(k_1) v_{s2}(k_2) \frac{\delta m (\omega)}{2}\, ,
\eea
where $\bar{u}_{s1}(k_1)$ and $v_{s2}(k_2)$ are the usual fermion-spinor factors. The rate of growth of the number density of the electron-positron pairs for a single $\omega$-mode can then be obtained from the above amplitude,\footnote{Again we can check against the perturbative result for an oscillating $\phi$, $\Gamma_{\phi \to ee}=g^2_{\phi e} m_\phi \beta_e^3/8 \pi$. }
 \bea
\dot{n}_{e^+e^-}(\omega)&=&\frac{ ( \delta m(\omega) \omega)^2  }{2}\int (2 \pi)^4 \delta(\omega-E_1-E_2)\delta^3(\mathbf{k_1}+\mathbf{k_2})\frac{d^3 k_1}{(2 \pi)^3 2E_1}  \frac{d^3 k_2}{(2 \pi)^3 2E_2}\nonumber
\eea
where the factor within the integral is just the 2-particle Lorentz Invariant phase space element, and upon integration, it gives,
\bea 
\dot{n}_{e^+e^-}(\omega)&=& \frac{(\delta m(\omega))^2   \omega^2  \beta_e^3}{16 \pi }.
\label{needot}
 \eea
Here $\beta_e=\sqrt{1-4 m_e^2/\omega^2}$ is the magnitude of the momentum of both the electron and the positron.

Some subtleties have not been considered in the above analysis. First, we have ignored that the electron and positron mass contributes from $\mu(\omega)$. Including this effect, the delta function involving the electron and positron energies implies
\bea
|\textbf{k}|^2=\frac{\omega^2}{4}-(m_e(1+\frac{\delta m (\omega)}{m_e}\cos \omega t))^2.
\eea
This gives an effective range, $k \in\omega/2\pm \delta k/2$, for the momenta, where,
\bea
\delta k\sim \frac{4\delta m(\omega) m_e}{\omega}.
\eea

The analysis leading to \eq{needot} also ignores the effect of the Pauli exclusion principle. To incorporate this effect, let us estimate the time for all the momentum eigenstates of the electron/positron to get occupied. We ignore the effect of the universe's expansion for a first estimate. The maximal number of electron-positron pairs that can be produced before the exclusion principle blocks the process can be computed by demanding that the occupation number for each momentum eigenstate is saturated. This gives,
\bea
{n}^{max}_{e^+e^-} &=&2 \frac{4 \pi}{(2 \pi)^3}\, k_\star^2  \mathop{\delta k}.
\label{nmax}
\eea
 We can thus estimate the time scale when particle production will stop because of the exclusion principle to be, 
\bea
\Delta t_{\rm pauli}=\frac{{n}^{max}_{e^+e^-}}{\dot{n}_{e^+e^-}}=\frac{16 m_e}{\pi \delta m(\omega)\omega}.
\label{pauli}
\eea
The expansion of the universe can suppress this effect by redshifting the momentum of the electron and positron. The time scale for this is, 
\bea
\Delta t_{\rm redshift}=\frac{\delta k}{k_\star}\frac{1}{H}=\frac{8 \delta m(\omega) m_e}{\omega^2  H}.
\label{redshift}
\eea
Pauli blocking is thus effective only if $\Delta t_{\rm pauli}<\Delta t_{\rm redshift}$.

For our final bounds, we require that the energy density injected in the $\Delta t=50$ years ($\Delta t=5$ Gyr\footnote{This corresponds to the time since dark energy-matter equality}) does not exceed the critical density; we show the excluded region in orange (hatched orange). For $\Delta t_{\rm pauli}<\Delta t_{\rm redshift}$, Pauli blocking is effective so that the above condition becomes,
\bea
\Delta \rho= \dot{n}_{e^+e^-}\omega \times \min(\Delta t_{\rm pauli},\Delta t) <\rho_c.
\label{wpauli}
\eea
On the other hand, for $\Delta t_{\rm pauli}>\Delta t_{\rm redshift}$,  we demand,
\bea
\Delta \rho= \dot{n}_{e^+e^-}\omega \Delta t <\rho_c.
\label{nopauli}
\eea
We find that the redshifting is relevant only for higher values of $\delta m (\omega)/m_e \gtrsim 10^{-18}-10^{-20}$, where the precise value depends on $\omega$. Pauli blocking is not effective in this region which gets excluded by \eq{nopauli}. For smaller values of $\delta m (\omega)/m_e$, \eq{wpauli} becomes relevant. For the lower boundary of the solid orange excluded region,  $\Delta t=50$ years is always less than $\Delta t_{\rm pauli}$. For the lower boundary of the hatched region $\Delta t_{\rm pauli}< \Delta t$ for $\omega<2\times 10^{11}$ eV and vice-versa for higher frequencies; this leads to the kink in the hatched region at this frequency value.

\subsubsection{$e\to e \gamma$ process}
\label{eeg3}
The region below the pair production threshold, $\omega < 2 m_e$, can be probed by the $e\to e \gamma$ process, which is the only probe in the range between 100 eV and $2 m_e$. The $\delta m (\omega)$ insertion now gives a Compton-like diagram (see Fig.~\ref{4figs}) that can be computed following Sect.~\ref{eviolation}. This agrees with the  $e \gamma \to e {\phi}$ amplitude in Ref~\cite{Grifols}. We again get the bounds shown in Fig.~\ref{gradient} by demanding that the ionisation time of hydrogen is less than 50 years (solid green) or the age of the universe (hatched green). A violation of energy conservation in the sun gives the hatched orange region. For details about the procedure to obtain these bounds, we refer to Sect.~\ref{eeg1}. Note, however, that for the bound arising from  VEC in the sun,  an essential difference from the
case discussed in Sect.~\ref{eeg1} is that there is no virtual photon here, so Debye screening effects are not relevant. The bound, in this case, extends to frequencies as low as the plasma mass of the photon in the sun $m_\gamma \approx 300$ eV, below which the process is again kinematically forbidden. 

\subsubsection{Electron emission due to photoelectric-like processes}
\label{pe1}
We want to study the emission of electrons from an atom by a process analogous to the photoelectric effect, with the TV couplings playing the role of the electromagnetic field. Following Ref~\cite{aelectric3}, the rate for this process can be derived by comparing the amplitude for this process with that for the photoelectric or the axioelectric effect~\cite{aelectric1,aelectric2, aelectric3}; the latter process takes place by the absorption of an axion instead of a photon. The amplitude for all three cases can be written in a similar form,
\bea
{\rm TV~due~to~\delta m:}&&\delta m (\omega)\langle E_f| e^{i\mathbf{k.r}}|E_i\rangle \approx i\delta m (\omega)\omega\langle E_f|\mathbf{v.r}|E_i\rangle\nonumber\\
{\rm Photoelectric~effect:}&&e A \omega \langle E_f|\mathbf{\epsilon.r}|E_i\rangle\nonumber\\
{\rm Axioelectric~effect:}&&g_{e\tilde{\phi}} \tilde{\phi} \frac{\omega^2}{2 m_e} \langle E_f|\mathbf{\sigma.r}|E_i\rangle
\label{3process}
\eea
where $\omega=E_f-E_i$. The same effect can also be induced by $\tilde {m}(\omega)$, and the amplitude, in this case, can be read off from the axioelectric case by replacing $g_{e\tilde{\phi}}\tilde{\phi} \to \tilde{m}$. The axioelectric effect has been probed by direct detection experiments such as CoGeNT~\cite{cogent}, CDMS~\cite{cdms}, EDELWEISS-II~\cite{edel} and XENON~\cite{xenon}. Amongst these, the most stringent bounds are from the XENON experiment. Using \eq{3process} we can recast these to obtain bounds on $\delta m(\omega)$  shown in Fig.~\ref{gradient}. 
\subsubsection{$e \gamma\to e \gamma$ process}
\label{egeg1}

The kinematical restrictions for the $e \to e \gamma$ process at low frequencies,  $\omega < m_\gamma$,  can be evaded if we consider the $e \gamma\to e \gamma$ process. There are six possible diagrams as shown in Fig.~\ref{4figs}; the three circles show the three possible positions for the TV insertion, and in each case, there is a $s$ and $u$ channel diagram. The TV insertion allows for an increase/decrease in energy. To compute the net effect, we compute the rate for the $e(p_1) \gamma(p_2)\to e(p_3) \gamma(p_4)$ process   and subtract the time reverse process. This yields,
\bea
\omega\int\prod_{i=1}^4 \frac{d^4 p_i}{(2 \pi)^3 2 E_i} |{ M}|^2 (f_{MB}(E_1) f_{BE}(E_2)-f_{MB}(E_3) f_{BE}(E_4)) \delta^4(\sum_{i=1}^4 p_i)
\eea
where the final state has greater energy than the initial state, i.e., $E_3+E_4=E_1+E_2+\omega$. The explicit form for the distribution functions, $f_{MB,BE}(E_i)$, are given in Sect.~\ref{eeg1}. The factor within parenthesis is always positive, i.e. the net effect is one of heating the sun. This is because the distribution functions,$f_{MB, BE}(E_i)$, fall with energy. We obtain the final bounds by requiring, as in Sect.~\ref{eeg1},  that the lumionosity of the sun does not increase by more than 10$\%$; these are shown in hatched brown in Fig.~\ref{gradient}. Let us now explain the shape of this curve.  At high frequencies, the diagrams with the TV insertion on the internal electron line contribute dominantly. The diagrams with the TV insertions on the external electrons start dominating for $\omega\lesssim E_\gamma\sim T$, where $T\sim$ keV is the solar core temperature. The contribution of these diagrams grows for small $\omega$ because of a factor  $1/(2 E \omega)$ in the electron propagator with the TV insertion, $E$ being the electron's energy. The bounds flatten with respect to $\omega$ in this region because the rate of the forward/backward processes that are proportional to $1/\omega^2$, the energy gained per scattering is $\omega$, and there is a further suppression by a factor, $\omega/T$. This would imply no weakening of the bounds even in the $\omega\to 0$ limit, which is likely to be unphysical.

A physical effect that cuts off this amplitude's IR growth is that the interaction time between the electron and photon cannot exceed the mean free time of the electron. The latter can be calculated from the collision frequency of the electron in the solar plasma,  $\Gamma_{\rm coll} \sim 1$ eV, which is given by standard expressions  (see, e.g., Ref.~\cite{lynn}). We regulate this by multiplying the squared amplitude by a somewhat ad hoc form factor,
\bea
\frac{\omega^2}{\omega^2+\Gamma_{\rm coll}^2}.
\label{ffac}
\eea
This form factor leads to the weakening of the bounds shown in Fig.~\ref{gradient}.
We again stress that this is a very rough and ad hoc approach. A better treatment would be desirable, but  is beyond the technical scope of the present work.

\subsection{Constraints on $\delta m$ from gradient forces and time  varying fundamental constant tests}

The constraints, both from gradient forces and those from the measurements of the time variation of fundamental constants, can be obtained entirely analogously to the discussion in Sect.~\ref{sec:delz}.
For the gradient forces, this is essentially done by applying Eq.~\eqref{delM}. For the various precision measurements, one has to determine the respective quantities' dependence on the electron mass variation. Indeed this is usually already done in the literature, and we use the results from there~\cite{gkr,wash,electronbbn,cmb,review,Hees:2016gop,DD,budkerN,antypas,budkerJ,Kennedy:2020bac,geo,holo,Grote:2019uvn,aint2,magis1,magis2,auriga,dmsound3,dmsound1}. The results are shown in Fig.~\ref{gradient}.

\begin{figure}
\noindent \begin{centering}
\hspace{-2em}\includegraphics[scale=0.48]{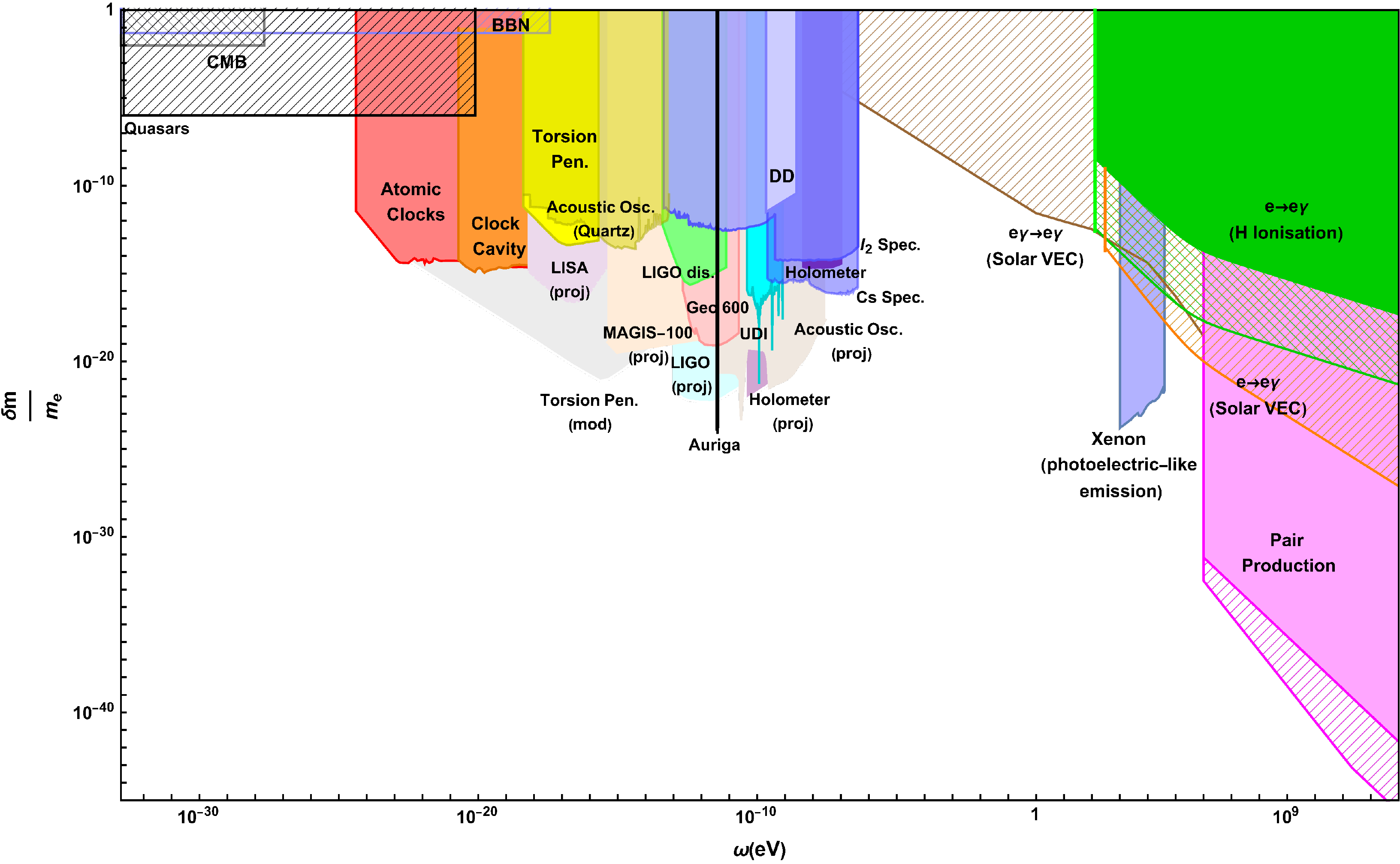}
\par\end{centering}
\protect\caption{Bounds on  $\delta{m}(\omega)$ from observational experimental and VEC probes. Bounds that only require the TV to be present in recent times (in the last 50 years) are shown in solid colours, whereas bounds that require us to assume TV throughout cosmological history are shown in hatched lines. See text for more details.}
\label{gradient}
\end{figure}

\section{Constraints on $\tilde{\cal m}(x)$}
\label{sec:tilm}
\subsection{Bounds from violation of energy conservation}
\subsubsection{Spontaneous production of electron-positron pairs}

Just as in Sect.~\ref{eepp},   the time dependence in  $\tilde {\cal m}(x)$ shown in \eq{defs} leads to spontaneous production of $e^+e^-$ pairs with the following amplitude, 
\bea
M_{0\to ee}&=&\langle \bar{\psi}(k_1,s_1) \gamma^5\psi(k_2,s_2) | \int d^4 x ~\tilde{\cal m}(x)~\bar{\psi}(x)\psi(x)|0\rangle
\nonumber\\&=&\sum_\omega (2 \pi)^4 \delta(\omega-E_1-E_2)\delta^3(\mathbf{k_1}+\mathbf{k_2})   \bar{u}_{s1}(k_1)\gamma^5  v_{s2}(k_2) \frac{\tilde{m} (\omega)}{2} \,.
\eea
We finally obtain the rate of growth of the number density of pairs,
  \bea
\dot{n}_{e^+e^-}(\omega)&=& \frac{ (\tilde{m}(\omega))^2 \omega^2}{16 \pi } \beta_e
\label{needot2}
 \eea
where $|\mathbf{k}|=\sqrt{\omega^2/4-m_e^2}$ is again the magnitude of the momentum of both the electron and the positron. 

We see that the expression in \eq{needot2} is identical to that in \eq{needot}  if we replace $\delta{m}(\omega) \to \tilde{m}(\omega)$  and neglect factors of $\beta_e$ which, in any case, is a unity to an excellent approximation apart from the small region in parameter space where $\omega \approx 2 m_e$. Thus the rest of the analysis of Sect.~\ref{eepp} can be carried over to this case. As in Sect.~\ref{eepp} the two time scales $\delta t_{\rm pauli}$ and  $\delta t_{\rm redshift}$ are given by \eq{pauli} and \eq{redshift} respectively with the above replacement. Again, in the region, Pauli blocking is relevant, i.e. for $\delta t_{\rm redshift}>\delta t_{\rm pauli}$, the particle production can last at most for the time period, $\delta t_{\rm pauli}$. Requiring again that the energy density injected in 50 years (5 Gyr) is less than the critical density, we get the excluded region shown in magenta (hatched magenta) in Fig.~\ref{mtilde}.

\subsubsection{$e\to e \gamma$ process}
\label{eeg4}
The $e\to e \gamma$ process is again the main probe between 100 eV and $2 m_e$. As before the $\tilde{m} (\omega)$ insertion gives an inverse Compton-like diagram (see Fig.~\ref{4figs}) that we compute following Sect.~\ref{eviolation}. This agrees with the $e \gamma \to e \tilde{\phi}$ amplitude in Ref~\cite{cpn}, adapted to our case. We show the bounds obtained in  Fig.~\ref{mtilde}; the bounds obtained by demanding that the ionisation time of hydrogen is less than the age of the universe (50 years) are shown in hatched green (solid green); the bounds from the VEC in the sun (hatched orange). For details about the procedure to obtain these bounds, we again refer to Sect.~\ref{eeg1}. As in Sect.~\ref{eeg3}, Debye screening is not relevant for this process, and the bounds extend to all frequencies larger than the photon plasma mass. 

\subsubsection{$e \gamma\to e \gamma$ process}
\label{egeg2}

As in Sect.~\ref{egeg1}, the $e \gamma\to e \gamma$ process does not have the kinematical limitations of the $e \to e \gamma$ process, which allows us to extend the bounds to lower frequencies. We follow the steps outlined in Sect.~\ref{egeg1} but now with a $\tilde{m} (\omega)$ insertion; this gives us the bounds shown in hatched brown in Fig.~\ref{mtilde}. 
We use the same ad hoc procedure as in Sect.~\ref{sec:delm} to regulate the infrared behaviour accordingly, the curve in Fig.~\ref{gradient} shows flattening below the temperature of the sun and a weakening of the bound below the collision frequency of the electron in the solar plasma where the latter effect again arises from the introduction of the form factor in \eq{ffac}.

\subsubsection{Emission of electrons due to photoelectric-like process}

The axioelectric bounds from  CoGeNT~\cite{cogent}, CDMS~\cite{cdms}, EDELWEISS-II~\cite{edel} and XENON~\cite{xenon} can be recast to obtain bound on $\tilde{m}(\omega)$ using \eq{3process}. We show the most stringent bounds from the XENON experiment~\cite{xenon}    in Fig.~\ref{mtilde}. 

\subsection{Experimental bounds on $\tilde{m}$.}

Using the alternate form for the lagrangian term in  \eq{alternate}, the     Hamiltonian can be written as,
\bea
H^{\tilde{ee}}=\frac{\nabla {\tilde{m}}. \sigma}{2m_e} +\dot{\tilde{m}}~\frac{ \mathbf{p.\sigma}}{2 m^2_e}
\label{ham}
\eea
in the non-relativistic limit. The above terms can be probed by different experimental techniques described below.

\paragraph{Electron Spin Resonance} The first term is \eq{ham} very similar to the interaction of an electron spin with an external magnetic field, with $\nabla {\tilde{m}}/m_e$ playing the role of the magnetic field. As first proposed in Ref.~\cite{barbieri}, this term can be probed by collective spin excitations in magnetic materials. The QUAX proposal~\cite{quax} aims to utilise this effect. We show bounds from the 2018~\cite{quaxm18} and 2019 runs~\cite{quaxm19} of the QUAX experiment and the results from Ref.~\cite{flower}  in Fig.~\ref{mtilde}. 

\paragraph{Atomic transitions} The second term can also lead to an electronic transition between two energy levels~\cite{Sikivie}. The energy difference between the levels can be tuned using the Zeeman effect. The  AXIOMA project~\cite{axioma1,axioma2} has begun feasibility studies to realise this idea. In the lighter shade of green, we show expected bounds from Ref.~\cite{Sikivie} in Fig.~\ref{mtilde}.

\paragraph{Polarised torsion pendulums}
Torsion pendulums can also probe the second term above with polarised electrons~\cite{grtorsion1, grtorsion2}. The existing bounds~\cite{TerranoSpin} on $\tilde{m} (\omega)$ are, however, too weak and not shown in Fig.~\ref{mtilde}. Note, however,  that if ALP dark matter induces an effective,  $\tilde{\cal m}={g}_{\tilde{\phi}e} \langle \tilde{\phi}\rangle$, interesting bounds on ${g}_{\tilde{\phi}e}$ can still be obtained at low masses because of a significant value for the oscillation amplitude, $\langle \tilde{\phi} \rangle=\sqrt{2 \rho_{DM}/m_{\tilde \phi}}$~\cite{TerranoSpin}.

\paragraph{Penning Traps}
Penning-trap experiments, like ATRAP~\cite{atrap} can measure cyclotron frequencies of a confined particle or antiparticle that interacts with electromagnetic fields.   The dominant effects arise from the interaction of the confined particle or antiparticle with the
constant magnetic field of the trap.\footnote{There is a quadrupole electric field to provide the axial confinement. This, however, only generates effects suppressed by
a factor of $E/B \sim 10^{-5}$.}  For $\omega\lesssim 10^{-23}$ eV, $\partial_\mu \tilde{m}(x)$ would effectively be constant over the time period of the ATRAP experiment where a single measurement took about an hour. Still, the final result was an average of nine measurements over a 6-month period~\cite{atrap}. In this regime the term, $ \partial_\mu \tilde{m}(x)  \bar{\psi} \gamma^\mu \gamma^5 \psi/2 m_e$ generates different shifts in the cyclotron frequency for $e^+$ and $e^-$. The theoretical expressions for these shifts have been computed in Ref.~\cite{Ding}. The ATRAP  experiment is sensitive to these differential shifts, which can be used to obtain the following bound:
\bea
\frac{\omega~\tilde{m}(\omega)}{m_e}\lesssim 10^{-8} {\rm eV}
\eea
which unfortunately does not result in a meaningful bound  on $\tilde{m}(\omega)$ in the relevant region, $\omega <10^{-23}$ eV. 
\begin{figure}
\noindent \begin{centering}
\hspace{-2em}\includegraphics[scale=0.55]{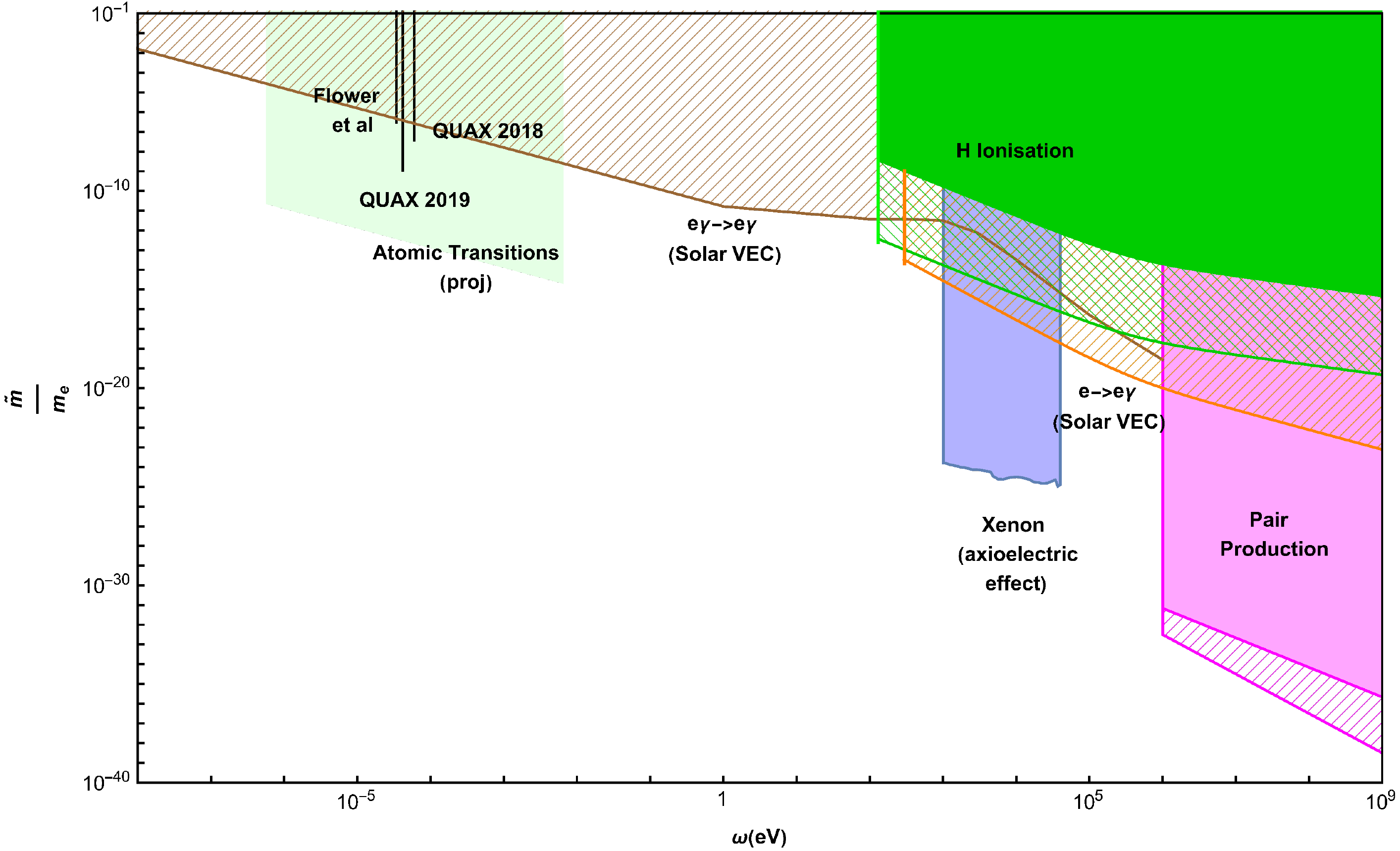}
\par\end{centering}
\protect\caption{Bounds on $\tilde{m} (\omega)$ from  VEC, experimental and observational  probes. Bounds that only require the TV to be present in recent times (in the last 50 years) are shown in solid colours, whereas hatched lines show bounds that require us to assume TV throughout cosmological history. See text for details.}
\label{mtilde}
\end{figure}

\section{Comparison with wave-like dark matter scenario}
\label{sec:compare}

In this section, we compare our results with previous literature on spacetime-dependent couplings. As we consider only breaking of time translation invariance, our results can be compared to models with a time variation of fundamental constants~\cite{review}--including those that aim to explain dark energy~\cite{de1,de2,de3}- or wavelike dark matter (WLDM) models~\cite{wdm1,wdm2} where such time variation arises dynamically from an underlying (pseudo-)scalar field. The first category of models involves slow time variation on scales of the order of the age of the universe, and thus the main probe for these is the comparison of the present values of the fundamental constants with inferred values from the past, such as the BBN, CMB quasar and Oklo bounds shown in Fig.~\ref{photon} and Fig.~\ref{gradient}. As far as the WLDM models are concerned,  the relevant frequency range, or equivalently the mass range, for the WLDM models is $10^{-22}$ eV $<\omega<10$ eV. We will discuss the lower bound on the frequency soon. The upper bound arises from requiring that the occupation number of dark matter quanta is large enough for it to be considered a classical field. As we discuss below, while some analogues of the VEC bounds discussed in our work exist even for these dynamical models, the previously discussed kinematical considerations push many VEC effects to frequencies higher than 10 eV.   VEC bounds have, therefore, usually not been discussed} in the literature on these dynamical models. In upcoming work, we will also consider the spatial variation of couplings which will lead to qualitatively very different probes from those for the above dynamical models.

For frequencies $\omega <10$ eV, there is, of course,  a lot of overlap between the probes of spacetime translation violation considered here and the observational/experimental probes already considered for the dynamical models. Even then, there are crucial differences between our scenario and dynamical models. These differences originate from the fact that in the dynamical scenarios, the observed effects arise from the expectation value of an actual dynamical field that has particle excitations, respects an equation of motion and carries energy and momentum that, in turn, have gravitational effects. We now discuss these differences in detail.

\paragraph{Absence of particle excitations} Some of the most potent constraints in the scenario with a dynamical (pseudo-)scalar, such as-- fifth force constraints, astrophysical bounds, helioscope bounds, bounds from beam dump experiments etc.-- are not relevant for TV probes as these arise from the exchange of the particle excitations. For instance, in Fig.~\ref{photon} and Fig.~\ref{gradient} if $\delta Z$ and $\delta m$ originated from a background scalar field, fifth force constraints will be the most powerful constraints for  $\omega\lesssim 10^{-24}$ eV or $\omega \gtrsim 10^{-16}$ eV. 
In the TV scenario, however, these constraints are absent, making $Cs$ and $I_2$ spectroscopy and experiments utilising interferometry and acoustic oscillations the most sensitive probes in the $10^{-16}$ eV $\lesssim \omega \lesssim 10^{-7}$ eV region. 

Similarly, suppose $\delta \tilde{Z}$ and $\delta \tilde{m}$ originated from a background pseudoscalar field. In that case, some of the most dominant bounds in different frequency ranges, such as -- helioscope bounds, astrophysical bounds from star cooling, the supernova SN1987a and cosmological bounds from a thermal population of the particles -- are absent in the TV case.

\paragraph{Absence of gravitational effects}  A second crucial distinction is that if the TV couplings in \eq{defs} do not arise from a (pseudo-)scalar field that carries energy density, there are no corresponding gravitational effects. This spacetime variation is thus not subject to standard constraints on dark matter or dark energy. In fact,  for the WLDM case,  dark matter masses $\lesssim$10$^{-22}$ eV are ruled out by Lyman-$\alpha$ constraints~\cite{irsic1,irsic2} and other astrophysical and cosmological considerations~\cite{marsh,schutz}. These arguments, however, do not apply to our TV couplings. Thus our plots can extend to frequencies as small as 10$^{-32}$ eV, which corresponds to the inverse of the age of the universe (we have not covered this full allowed range of frequencies in  Figs.~\ref{photon2} and ~\ref{mtilde} for presentation purposes).

\paragraph{Difference in cosmological evolution}  Another essential difference is that in the WLDM scenario, the amplitude for the oscillations cannot be assumed to be constant as in the TV scenario. This is because the underlying scalar/ field will experience a damping $\phi \sim a^{-3/2}$ as the universe expands. This results in a difference in the bounds, which depend on cosmological history (hatched in our plots). This includes bounds which depend on the difference in the time-dependent couplings between the present moment and a past time. Such bounds will be much stronger in the dark matter scenario owing to the larger amplitude in the past. For instance,  cosmic birefringence implies strong bounds in the DM scenario~\cite{Trivedi} but weak constraints in the TV scenario. This fact also makes an important distinction for the spontaneous particle production bounds. This effect, interpreted as a decay of the condensate in the coherent dark matter scenario, results in a stronger bound at earlier times once the $a^{-3/2}$ scaling of the oscillation amplitude is taken into account (see Ref.~\cite{surprise}). In contrast, the best bounds typically arise today in the TV violation scenario (we assume that the amplitude and frequency of the TV oscillation are constant).

\paragraph{Absence of backreaction from particle production} For the WLDM case, there is an important difference between its decay to photon or electron pairs and spontaneous particle production.\footnote{The relevant frequencies for models with a slow variation of fundamental constants are far below the pair production thresholds.} In the WLDM case, there is a backreaction on the DM. The energy density of the dark matter and, thus, its oscillation amplitude must get depleted. There is no such backreaction or depletion of the couplings in our scenario, as now we have a genuine violation of energy conservation. Thus one of the conditions used to derive our spontaneous particle production bounds -- that the energy density of the decay products should not exceed the critical density -- would not be relevant in the dark matter scenario. This is because the energy density of the decay products would, in any case, never exceed the dark matter energy density by conservation of energy. Instead, the bound in the dark matter case would arise from requiring no appreciable depletion of dark matter~\cite{surprise}.  

\paragraph{Absence of inhomogeneties and differences in frequency spread}
Another critical difference in the WLDM  case is the formation of halos. The energy density of WLDM near Earth would be many orders of magnitude higher than its average density in the universe. This is one of the reasons that the particle production bounds in the dark matter scenario are relatively less sensitive when compared to other probes, such as haloscopes or experiments testing the variation of fundamental constants ~\cite{surprise}. In the TV case, on the other hand, these probes are closer in sensitivity as the TV  couplings have been assumed to be spatially constant. Even if we relax our assumption of spatial uniformity, there is no reason to expect that the spatial inhomogeneities in the TV couplings will be correlated to gravitational dynamics.

The velocity spread of virialized dark matter can also be determined and implies a typical frequency spread $\Delta \omega/\omega \sim 10^{-6}$~\cite{redondo}. This frequency spread can actually be easily resolved in haloscopes~\cite{redondo}. For models involving slow variation of fundamental constants, it is again not expected that only a  single frequency would be present. That said, our simplifying assumption of having one sharp frequency mode will provide a much narrower signal than these scenarios. If we go beyond our assumption of a single-mode being turned on, the frequency dependence of the four  TV couplings, $\alpha_i^{\rm TV}(\omega)$, in \eq{defs} would be arbitrary and in general not match the DM frequency spread $\Delta \omega/\omega \sim 10^{-6}$.

\section{Conclusions}
\label{sec:conclusion}

In this paper, we have taken the first step of putting one of the most fundamental assumptions of physics to a test: the independence of fundamental physical laws concerning time translation. This is an essential part of the larger Poincare symmetry and, of course, the origin of energy conservation.
While this is deeply ingrained in physics, putting it to strict and highly precise tests is nevertheless non-trivial.
To have a concrete realisation, we needed to make assumptions on how the fundamental laws would now explicitly depend on time. For the present paper, we chose a simple, periodic variation and specified the limits in frequency and amplitude for various couplings of the electromagnetic sector. This simple choice allowed us to adopt a wide range of existing experimental results, particularly from searches for ``wave-like'' bosonic dark matter. We stress, however, that despite similarities to the dark matter scenario, there are potentially also important differences because such a time translation violation will not necessarily be bound by the same constraints as dark matter, such as fulfilling the energy density requirement, cosmological evolution, gravitational clumping etc., discussed in more detail in Sect.~\ref{sec:compare}.

We have considered TV effects in couplings of electrons and photons. We have shown that at the level of operators with dimension $\leq 4$, a general parametrisation of such effects can be done with the four couplings, $\delta {\cal Z}(x), \tilde{\cal Z}(x), \delta {\cal m}(x),\tilde{\cal m}(x)$ in \eq{lagr}. These couplings have been assumed to have temporal but no spatial variation in the FRW frame (see \eq{defs}). The three main kinds of constraints on these couplings arise from (1) observational and experimental probes of this time variation,   (2) effects, such as gradient forces, due to the spatial variation of the couplings in the earth frame of reference and (3) conservation of energy violating scattering processes such as spontaneous pair production of electrons and photons from the vacuum, the $e\to e \gamma$ process etc.. Our main result is the constraints on these couplings presented in Fig.~\ref{photon}-\ref{mtilde}.

Our present investigation is simplistic in two important ways.
The first is our focus on the time translation invariance. 
An obvious path is considering spatial variations and the corresponding momentum conservation. Second, we have taken a specific single-frequency violation of the time translation invariance. Improving this may be non-trivial. One way would be to consider alternative spectra of violations, such as e.g. a white noise form. More generally, one can also hope to set a limit at the spectral density such that limits on any spectral form of the violation can be obtained by integrating the violation ``signal'' with the limiting ``noise'' curve. 
Building on the current setup and developing a framework to coherently investigate general violations of Poincare invariance (much akin to the Standard Model Extension for Lorentz invariance violations~\cite{Colladay:1998fq}) seems a logical next step for a thorough investigation of the experimental and observational foundations of Poincare invariance.


\bigskip

\section*{Acknowledgements}
We would like to thank Dima Budker for discussions. JJ acknowledges support from the EU via the ITN HIDDEN (No 860881) and would like to thank the IPPP for a DIVA fellowship. RSG would also like to thank the IPPP for a DIVA fellowship.


\bibliographystyle{utphys}
\bibliography{references}

\end{document}